**Hopping Transport through Defect-induced Localized States in Molybdenum Disulfide**

Hao Qiu[1], Tao Xu[2], Zilu Wang[3], Wei Ren[4], Haiyan Nan[3], Zhenhua Ni[3], Qian Chen[3], Shijun Yuan[3], Feng Miao[5], Fengqi Song[5], Gen Long[1], Yi Shi[1,*], Litao Sun[2,*], Jinlan Wang[3,*] and Xinran Wang[1,*]

[1]*National Laboratory of Microstructures and School of Electronic Science and Engineering, National Center of Microstructures and Quantum Manipulation, Nanjing University, Nanjing 210093, P. R. China*

[2]*SEU-FEI Nano-Pico Center, Key Laboratory of MEMS of Ministry of Education, Southeast University, Nanjing, 210096, P. R. China*

[3]*Department of Physics, Southeast University, Nanjing 211189, P. R. China*

[4]*Department of Physics, Shanghai University, 99 Shangda Road, Shanghai 200444, P. R. China*

[5]*School of Physics, Nanjing University, Nanjing 210093, P. R. China*

* Correspondence should be addressed to X. W. (xrwang@nju.edu.cn), J. W. (jlwang@seu.edu.cn), L. S. (slt@seu.edu.cn) and Y. S. (yshi@nju.edu.cn).

## Abstract

Molybdenum disulfide is a novel two-dimensional semiconductor with potential applications in electronic and optoelectronic devices. However, the nature of charge transport in back-gated devices still remains elusive as they show much lower mobility than theoretical calculations and native n-type doping. Here we report transport study in few-layer molybdenum disulfide, together with transmission electron microscopy and density functional theory. We provide direct evidence that sulfur vacancies exist in molybdenum disulfide, introducing localized donor states inside the bandgap. Under low carrier densities, the transport exhibits nearest-neighbor hopping at high temperatures and variable-range hopping at low temperatures, which can be well explained under Mott formalism. We suggest that the low-carrier-density transport is dominated by hopping via these localized gap states. Our study reveals the important role of short-range surface defects in tailoring the properties and device applications of molybdenum disulfide.

**Introduction**

Layered transition metal dichalcogenide is an emerging class of two-dimensional (2D) materials with attractive electronic and optical properties[1-3]. Among them, molybdenum disulfide ($MoS_2$) is a semiconductor with bandgap of ~1.1-2eV (Ref. 4, 5), which is more suitable than graphene for CMOS-like logic device applications. The true 2D nature makes $MoS_2$ outperform Si transistors at the scaling limit[6,7]. Recently, many $MoS_2$-based devices have been demonstrated, including field-effect transistors[8-10], integrated circuits[11] and photo-transistors[12]. Gate-tunable metal-insulator transition and superconductivity have been observed in $MoS_2$ (Ref. 13).

Despite the rapid progress in $MoS_2$ device applications, the nature of charge transport still remains elusive. In particular, the mobility of back-gated few-layer $MoS_2$ (meaning 1-3 layers in the rest of the paper) devices was usually over an order of magnitude lower than the theoretical limit set by phonon scattering[8,10,11,14,15]. These devices exhibit insulating behavior at low carrier densities[10,13,16], the microscopic understanding of metal-insulator transition is still lacking[13]. Several possible scattering mechanisms have been proposed including phonons[15,17], Coulomb impurities[8,18] and short range disorder such as interfacial bonding/roughness[19], yet the nature of the disorders remains to be unequivocally identified. Recently, the role of dielectric screening was also heavily debated[14]. On the other hand, the conductivity of few-layer $MoS_2$ is extremely sensitive to ambient environments[10, 20], indicating an important role of surface defects in charge transport. In addition, the origin of n-type doing in few-layer $MoS_2$ devices is still unclear.

In this work, we combine variable-temperature transport measurements, aberration-corrected TEM, density functional theory (DFT) and tight-binding calculations to investigate the nature of charge transport in few-layer $MoS_2$. We observe a high density (~$10^{13}cm^{-2}$) of sulfur vacancies (SV) on the surface of $MoS_2$ which act like electron donors and induce localized states in the bandgap. Below a critical carrier density, electrons are localized, and transport is dominated by nearest-neighbor hopping (NNH) and variable-range hopping (VRH) at high and low temperatures respectively. The experimental data can be well explained under Mott's formalism. Our model suggests a microscopic picture of metal-insulator transition and n-type doping observed in $MoS_2$ samples. Finally, we discuss the effect of screening and thickness on the charge transport of $MoS_2$.

**Results**

**Device fabrication and transport measurement**

The back-gated few-layer $MoS_2$ devices were fabricated on 300nm $SiO_2$/Si substrate as described in our previous work[10] (see Methods). We have measured 15 samples, all of which showed the same qualitative behavior. We present the data from a representative single-layer sample and bi-layer sample in the main text and supplementary material respectively. Fig. 1a shows the room temperature transfer ($I_{ds}$-$V_g$) characteristics of the single-layer $MoS_2$ device after vacuum annealing at 350K. The single-layer nature was confirmed by the position and relative intensity of the zone-center $E_{2g}^1$ mode (385.9$cm^{-1}$) and $A_{1g}$ mode (404$cm^{-1}$) in Raman spectroscopy[21] (Fig. 1b). The device exhibited an n-type transistor behavior with on/off ratio over $10^8$ and a field-effect mobility of 7.7$cm^2V^{-1}s^{-1}$, which was comparable to the mobility of back-gated devices made similarly. The mobility of the same device measured in ambient was 0.57$cm^2V^{-1}s^{-1}$ due to oxygen and water absorption[10].

We measured the conductivity as a function of carrier density $n=C_g(V_g-V_{th})$ in a cryogenic probe station from room temperature down to ~20K, where $C_g$=1.15×10$^{-4}$ Fm$^{-2}$ is the gate capacitance and $V_{th}$ is the threshold voltage extracted by linear extrapolation of $I_{ds}$-$V_{ds}$ characteristics in the linear region (Fig. 1a). We focused our attention on low density regimes up to ~3×10$^{12}$cm$^{-2}$, much lower than that achieved using ionic liquid gate[9,13]. Insulating behavior was observed throughout the whole temperature and carrier density range[16], which ruled out phonons as the dominant scattering source. Fig. 2b shows the Arrhenius plot of the normalized conductivity $\sigma$ under various $n$. The charge transport exhibited two distinct scaling behavior at high and low temperature regimes, separated by a characteristic temperature $T^*$ on the order of 100K. At $T>T^*$, the transport was activated, which could not be explained by Schottky barrier at the contacts but rather came from the channel[16], as $T_0$ was an order of magnitude smaller than the expected barrier height (>0.33eV) considering the work function of Ti contacts and electron affinity of MoS$_2$ (Ref. 22). At $T<T^*$, $\sigma$ transformed to a much weaker temperature dependence that could be fitted very well by the 2D VRH equation

$$\sigma \sim \exp\left[-\left(\frac{T_1}{T}\right)^{1/3}\right]. \qquad (1)$$

Such characteristic temperature dependence has been observed in many low dimensional systems and is a signature of hopping transport[30,31].

The localized states and low mobility are not likely introduced by long-range Coulomb impurities from the SiO$_2$ substrate. The mobility of graphene on similar SiO$_2$ substrates is limited by Coulomb effect to be ~10$^4$cm$^2$V$^{-1}$s$^{-1}$ (Ref. 23, 24), much higher than the intrinsic MoS$_2$ mobility limit. Another possibility is the chemical bonding to SiO$_2$ (Ref. 19), which lacks direct evidence. In addition, such chemical bonding would require dangling bonds in MoS$_2$, most likely associated with structural defects. In light of the TEM characterizations below, we attribute the localized states to short-range structural disorders in $MoS_2$.

**TEM characterization**

To reveal nature of the disorders, we transferred the exfoliated $MoS_2$ samples to TEM grids and performed aberration-corrected TEM characterizations under 80keV acceleration voltage (see Methods). A typical atomic image of a single-layer $MoS_2$ showed the expected hexagonal symmetry, but with a high density of defects as bright spots in position of sulfur atoms (Fig. 3a, red arrows). The defects can be clearly distinguished by analyzing the intensity profile. Comparison between intensity profiles along the red dashed line in Fig. 3a and simulated HRTEM image of single SV[25] shows quantitative agreement (Fig. 3b), which confirms that most defects are single SV. Statistical analysis on more than 30 areas from 3 samples showed the density of SV on the order of $10^{13}cm^{-2}$ (Fig. 3c), corresponding to an average defect distance $a$~1.7nm. We also observed other point or line defects, but with much lower density.

Great care was taken in TEM experiments to prevent knock-on damage or lattice reconstruction caused by 80keV electron beam irradiation[26, 27]. To rule out the effect of electron irradiation on vacancy density, we studied the evolution of defects under different exposure time. As shown in Supplementary Figure S2, we did not observe significant number of new SV after 44s exposure. The vacancy number increased from 65 to 70 in a $12 \times 12$ nm monolayer $MoS_2$ area after 62s exposure. The average vacancy generation rate was $5.6 \times 10^{10}cm^{-2}s^{-1}$, which was in a reasonable agreement with the value of $3.8 \times 10^{10}cm^{-2}s^{-1}$ in Ref. 27. The estimate of vacancy density introduced by electron irradiation during imaging was ~$10^{12}cm^{-2}$, which was far less than the statistical results shown in Fig. 3c. We thus concluded that the SVs were intrinsic rather than introduced by electron beam irradiation.

After hundreds of seconds of irradiation, we started to observe line defects, nanopores or even non-stoichiometric structure in $MoS_2$ (see Supplementary Figure S3), but those defects were irrelevant to the current work.

**DFT calculation**

The SV causes unsaturated electrons in the surrounding Mo atoms and acts as electron donors to make the $MoS_2$ electron rich as experimentally observed[8, 10, 11, 13, 16]. To understand the implication of SV to the band structure of $MoS_2$, we performed the DFT calculations on a 5×5 single-layer $MoS_2$ supercell with a single SV (Fig. 4a, see Methods), corresponding to a SV density of ~$4.6 \times 10^{13} cm^{-2}$. The defect-free $MoS_2$ has a direct bandgap of 1.67eV, consistent with earlier DFT calculations[28] and experiments[4, 5]. The presence of the SV introduces defect states in the bandgap and leads to a transition from direct to indirect 1.61eV bandgap. The gap state is a deep donor state (0.46eV below conduction band minimum) and is composed of two individual states, which are degenerate at Γ point but separated by 14meV at M point (Fig.4a). The bottom of conduction band is dominated by Mo 4d orbitals, showing a strong degree of delocalization (see Supplementary Figure S4). On the contrary, the gap state and top of valence band are the results of hybridization between strong Mo 4d orbitals and weak S 3p orbitals. The nearly dispersionless midgap state indicates a high electron mass and strong localization near the five-fold Mo atoms surrounding the SV, which is evident from the spatial mapping of these states (Fig. 4b inset). The electrons are mainly distributed within 3Å radius surrounding the SV (Fig. 4b). In contrast, electrons in the valance band of perfect $MoS_2$ are delocalized.

We can integrate the electron density of the localized midgap state (Fig. 4b) to obtain the total charge of 3e donated by a single SV, which, as expected, come from the three unsaturated Mo atoms. When $n$ is below a critical density ($\sim 10^{13}$-$10^{14}$cm$^{-2}$), the Fermi level is in the vicinity of gap states, electrons are localized. As $n$ is above the critical density, the gap states are filled, and band-like transport is expected (Fig. 2a). This picture naturally explains the metal-insulator transition observed in MoS$_2$ devices[13, 42]. We also performed DFT calculations on smaller supercells (4×4, 3×3, and 2×2) corresponding to higher vacancy concentrations. As shown in Supplementary Figure S5-S7, the defect states are broadened and have a tendency to be more delocalized with the increase of the SV concentration. This is a result of increased interactions between sulfur vacancies in adjacent supercells. We note that the threshold carrier density for the metal-insulator transition in thick ($\sim$20nm) sample[13] was about an order of magnitude lower than predicted by our model, likely due to screening effects and variations of defect density. Evidence of midgap states in MoS$_2$ was also reported previously[36].

**Discussion**

The above analysis makes it logical to interpret the transport data at low carrier densities by hopping mechanism (Fig. 2a), which has been widely observed in disordered materials[29-31]. At high temperatures, the transport is dominated by NNH showing an activated behavior $s \sim P \sim \exp\left(-T_0/T\right)$ with $T_0 \sim$ 300-650K (Fig. 2b, c). Following Mott's approach[32], assuming a constant density of states (DOS) $N_\mu$ near the Fermi level, the activation barrier $k_B T_0 = \left(N_m a^2\right)^{-1}$ is the energy spacing between nearest sites. Taking $T_0$=500K and $a$=1.7nm, $N_m = 0.8 \times 10^{15}\, eV^{-1} cm^{-2}$, within the range of DFT calculations (Considering Gauss broadening, $N_m = 58 eV^{-1} * 4.6 \times 10^{13}\, cm^{-2} = 2.67 \times 10^{15}\, eV^{-1} cm^{-2}$ for the midgap state in Fig. 4a). $T_0$ was found to decrease with increasing $n$ (Fig. 2c). Such trend could be explained by the increase of $N_m$ as the Fermi level approaches the mobility edge.

As $T$ decreases, hops over longer distance but with closer energy spacing are more favorable and the transport changes to Mott VRH[32]. According to Ref. 29, for 2D systems, the electron hopping probability

$$P \sim \exp\left(-2R/x - \frac{1}{pR^2 N_m k_B T}\right) = \exp\left(-2R/x - \frac{T_0}{p(R/a)^2 T}\right), \quad (2)$$

where $x$ is localization length. According to Fig. 4b, most of the localized electrons are distributed within 6Å diameter regime surrounding a SV. To further confirm the length scale of $x$, we employed tight binding calculations and Anderson model (see Methods). Supplementary Figure S9 gives the correlation between $x$ and average conductance $G$, which is dependent on the Anderson disorder parameter $W$ describing the random potential in the sample (see Supplementary Figure S8). The average conductance was obtained over an ensemble of samples with different configurations of the same disorder strength $W$. In the localized regime (the low conductance regime), $x$ was on the order of few angstroms, which was consistent with the distribution of electron wave function in the DFT calculations. We note that the slight variation of $x$ does not affect our conclusions qualitatively, so we use $x = 6$Å in the following discussions.

The optimal hopping distance $R_0 = \left(\dfrac{T_0}{pT} a^2 x\right)^{1/3} \equiv \dfrac{x}{3}\left(\dfrac{T_1}{T}\right)^{1/3}$ can be obtained by minimizing the right hand side of Eq. (2), where

$$T_1 = \frac{27 T_0}{p}\left(\frac{a}{x}\right)^2 \quad (3)$$

is the characteristic temperature in Eq. 1. The cross-over temperature $T^*$ is naturally given by $R_0=a$ (Ref. 31), which yields $T^* = \dfrac{T_0}{p(a/x)} \approx 0.11T_0$. The experimental data

is consistent with the hopping model at low electron densities ($T^*\sim0.15T_0$ at

$n=10^{11}\mathrm{cm}^{-2}$). To account for the discrepancy at high densities ($T^*\sim0.35T_0$ at

$n=2.5\mathrm{x}10^{12}\mathrm{cm}^{-2}$), we estimate the Debye screening length $l_D\sim1.1\mathrm{nm}$ at 100K, which

becomes comparable to $x$. This results in an increase of $x$ due to enhanced screening[33] and accordingly, $T^*/T_0$.

Next we discuss $T_1$ and its dependence on carrier density and dielectric

environment. For the back-gated device, $T_1/T_0 \sim90$ at $n=10^{11}\mathrm{cm}^{-2}$, and gradually

decreased to $\sim20$ at $n=2.5\times10^{12}\mathrm{cm}^{-2}$ (Fig. 2d). The hopping model in Eq. 3 predicts

$T_1/T_0=69$, which again agrees well with experiments at low carrier densities. The

decrease of $T_1$ at high densities could also be explained by enhanced screening[33]. In fact, when the Debye screening length is smaller than $x$ (at higher $n$), electrons are delocalized, the hopping model is no longer valid. To further test the hopping model, we spin coated PMMA (dielectric constant $\sim3.7$) after the low temperature measurements and immediately cool down the device again (see Supplementary while $T_0$ did not change significantly (Fig. 2c, d). Due to the change of dielectric environment, $l_D$ was expected to increase by a factor of $\sim2$. Such screening effect led to a more localized trapping potential and higher $T_1$ accordingly.

Figure S10). We observed an increase of $T_1$ by a factor of $\sim5$ after PMMA coating,

We also investigated thicker samples up to tri-layer, all of which showed hopping transport behavior. In Supplementary Figure S11, data from a bi-layer sample is presented. We observed variations of the characteristic temperatures for samples with different thickness. Such variations could be due to sample quality variation, screening by adjacent layers[19] or interlayer coupling[34, 35]. For thicker samples, we expect that the enhanced screening leads to a higher $T^*/T_0$. In order to test such prediction, we fabricated a device on a 4.9nm thick sample, corresponding to ~7 layers. The electrical data was summarized in Supplementary Figure S12. For $n$=1-2.5×$10^{12}$cm$^{-2}$, $T^*/T_0$ is in the range of 0.29-0.54, indeed higher than the single-layer device (Fig. 2). We also observed metallic behavior above ~250K, indicating the increasing contribution of bulk transport[17]. As the sample thickness further increases, bulk contribution becomes dominant[19], and at some point, a breakdown of the 2D hopping model is expected. In fact, the higher mobility in thicker devices indeed suggested carrier scattering dominated by phonons and long range Coulomb impurities[17,19].

In conclusion, we showed that the charge transport of few-layer MoS$_2$ in low carrier density regime can be explained by hopping through defect-induced localized states. The hopping model was supported by the direct TEM observation of SV, DFT calculations of band structure, and tight-binding calculations of localization length. Our study revealed the important role of short-range disorders on the MoS$_2$ device performance and the need to further improve sample quality by methods such as chemical vapor deposition[43, 44].

## Methods

### MoS$_2$ device fabrication and measurement

We did micro-cleavage of MoS$_2$ flakes (SPI supplies) on degenerately doped Si substrates with 300nm thermal oxide. We used photolithography to pattern source/drain contacts, followed by electron-beam evaporation of 40-nm-thick Ti electrodes and lift-off. Ti contacts were used to give low contact resistances[37]. The devices were further annealed at 300°C in a mixture of hydrogen and argon to improve contacts. Electrical measurements were carried out in a close-cycle cryogenic probe station with base pressure ~$10^{-5}$ Torr.

**TEM characterizations of MoS$_2$ samples**

MoS$_2$ samples were exfoliated by micromechanical cleavage techniques from natural MoS$_2$ bulk crystals (SPI supplies), and transferred to a holy carbon coated copper TEM grid by PMMA based transfer method[38]. In order to remove the polymer residue on the sample, annealing at 200°C in a mixture of hydrogen and argon was carried out before TEM.

TEM imaging was carried out in an image aberration-corrected TEM (FEI Titan 80-300 operating at 80 kV), and a charge-coupled device camera (2k × 2k, Gatan UltraScanTM 1000) is used for image recording with an exposure time of 1 s. The third order spherical aberration was set in the range 1-6μm, and the TEM images were recorded under slightly underfocused. Using a positive value of the third-order spherical aberration, a small defocus yields black atom positions in the TEM images. The electron beam was set to normal imaging conditions with uniform illumination at beam current density ~ 3Acm$^{-2}$. To enhance the visibility of the defects, a Fourier filter was applied to all TEM images. The atomic defects were also visible in the original images, although with lower signal-to-noise ratio (see Supplementary Figure S1). It is noteworthy that long time electron irradiation at 80 kV could generate sulfur vacancies due to knock-on damage[26, 27]. Therefore after we moved to an area, we limited the exposure time to less than 30s (to adjust focus) before imaging.

**Details of DFT calculations**

The DFT calculations were carried out using the Vienna *ab* initio simulation package (VASP)[39] with projected-augmented-wave pseudopotentials[40] and the Perdew-Burke-Ernzerhof (PBE) exchange and correlation functional[41]. The kinetic energy cutoff for plane wave basis set is 400eV. All atomic positions were fully relaxed without any symmetry constraint until the Hellmann–Feynman force on each ion and total energy change are less than 0.01eVÅ$^{-1}$ and $1\times10^{-4}$eV, respectively. The Brillouin zone is sampled by $5\times5\times1$ and $9\times9\times1$ k-point meshes within Monkhorst-Pack scheme[41] for geometry optimizations and density of states, respectively.

**Details of tight-binding calculations**

The tight binding method was applied to calculate electronic transport and estimate the localization length $x$ of $MoS_2$ monolayer. The calculation involves the defect structure created by the SV and a random on-site potential $W$ in the spirit of Anderson model. The disordered system has $5\times5$ supercells (see Supplementary Figure S4). We have adopted the following empirical parameters (all in unit of eV): hopping integral energy $t$=0.5 describes the bonding between neighboring Mo and S atoms, on-site potentials of +1 and -1 correspond to the Mo cation and S anion respectively, on top of that a random on-site potential with a uniform distribution [$-W/2,W/2$] is added all lattice sites. For monolayer $MoS_2$ with $W$=0 and no SV, the system exhibits a band gap of 2eV, close to experimental[4] and other theoretical results[45].

In Anderson model, $x$ depends on the random on-site potential $W$, which is difficult to measure experimentally. In order to correlate more with experiment, we calculated the conductance of the system under different $W$. We placed two semi-infinite metallic leads on both ends of the $MoS_2$ sample and set the on-site energy of the leads to zero to mimic metallic electrodes bridged by the central $MoS_2$. The conductance of the system is calculated by using the Green's function and Landauer formalism[46, 47]. As expected, we observed the evolution of conductance by varying the disorder strength $W$, from ballistic, diffusive to the localized regimes (see Supplementary Figure S8). Finally, $x$ could be defined as $x = L/[\ln(N_c) - \langle\ln(G)\rangle]$, where $L$ and $N_c$ are the sample length and the number of conducting channel respectively. An increased sample size (like the ones used in experiments) may give much larger conductance at higher energy levels, but such size dependence does not exist for electron transport at the Fermi level which is relevant to our moderate gating voltage.

**Acknowledgements.** This work was supported in part by Chinese National Key Fundamental Research Project 2013CBA01600, 2011CB922100, 2010CB923401, 2011CB302004, National Natural Science Foundation of China 61261160499, 11274154, 21173040, 11274222, 21373045, National Science and Technology Major Project 2011ZX02707, Natural Science Foundation of Jiangsu Province BK2012302, BK2012322, Specialized Research Fund for the Doctoral Program of Higher Education 20120091110028, Shanghai Supercomputer Center, Shanghai Shuguang Program 12SG34, and Eastern Scholar Program from the Shanghai Municipal Education Commission.

**Author Contributions.** H. Q., T. X and Z. W. contributed equally to this work. X. W., J. W., L. S. and Y. S. conceived the project. H. Q., X. W., F. M., F. S., and G. L. carried out device fabrication, electrical measurements and data analysis. T. X. and L. S. carried out TEM characterizations and analysis. Z. W., W. R., Q. C., S. Y. and J. W. carried out DFT and tight-binding calculations. X. W., J. W. and L. S. co-wrote the paper with all authors contributed to the discussion and preparation of the manuscript. Correspondence should be addressed to X. W. (xrwang@nju.edu.cn) and Y. S.

(yshi@nju.edu.cn) for general aspects of the paper, to J. W. (jlwang@seu.edu.cn) for

DFT calculations, and to L. S. (slt@seu.edu.cn) for TEM characterizations.

**Competing financial interests statement.** The authors declare no competing financial interest.

**References:**

1    Q. H. Wang et al., Electronics and optoelectronics of two-dimensional transition metal dichalcogenides, Nature Nanotech. 7, 669 (2012).

2    M. Xu, T. Liang, M. Shi & H. Chen, Graphene-like two dimensional materials, Chem. Rev. (2013) **DOI:** 10.1021/cr300263a.

3    X. Song, J. Hu & H. Zeng, Two-dimensional semiconductors: recent progress and future perspectives, J. Mat. Chem. C, (2013) **DOI:** 10.1039/C3TC00710C.

4    K. F. Mak, C. Lee, J. Hone, J. Shan & T. F. Heinz, Atomically thin $MoS_2$: a new direct-gap semiconductor, Phys. Rev. Lett. 105, 136805 (2010).

5    A. Splendiani, L. Sun, Y. B. Zhang, T. S. Li, J. Kim, C.-Y. Chim, G. Galli & F. Wang, Emerging photoluminescence in monolayer $MoS_2$, Nano Lett. **10**, 1271 (2010).

6    Y. K. Yoon, K. Ganapathi & S. Salahuddin, How good can monolayer $MoS_2$ transistors be? Nano Lett. **11**, 3768 (2011).

7    L. Liu, S. B. Kumar, Y. Ouyang & J. Guo, Performance limits of monolayer transition metal dichalcogenide transistors, IEEE. Trans. on Electron Devices **58**, 3042 (2011).

8    B. Radisavljevic, A. Radenovic, J. Brivio, V. Giacometti & A. Kis, Single-layer $MoS_2$ transistors, Nature Nanotech. **6**, 147 (2010).

9    Y. Zhang, J. Ye, Y. Matsuhashi & Y. Iwasa, Ambipolar $MoS_2$ thin flake transistors, Nano Lett. (2012).

10   H. Qiu, L. Pan, Z. Yao, J. Li, Y. Shi & X. Wang, Electrical characterization of back-gated bi-layer $MoS_2$field-effect transistors and the effect of ambient on their performances, Appl. Phys. Lett. 100, 123104 (2012).

11   H. Wang et al., Integrated circuits based on bilayer $MoS_2$ transistors, Nano Lett. 12, 4674 (2012).

12   Z. Yin et al., Single-layer $MoS_2$ phototransistors, ACS Nano 6, 74 (2012).

13   J. T. Ye, Y. J. Zhang, R. Akashi, M. S. Bahramy, R. Arita & Y. Iwasa, Superconducting dome in a gate-tuned band insulator, Science 338, 1193 (2012).

14   M. S. Fuhrer & J. Hone, Measurement of mobility in dual-gated $MoS_2$ transistors, Nature Nanotech. 8, 146 (2013).

15   K. Kaasbjerg, K. S. Thygesen & J. W. Jacobsen, Phonon-limited mobility in n-type single-layer $MoS_2$ from first principles, Phys. Rev. B 85, 115317

(2012).

16    A. Ayari, E. Cobas, O. Ogundadegbe & M. S. Fuhrer, Realization and electrical characterization of ultrathin crystals of layered transition-metal dichalcogenides, J. Appl. Phys. 101, 014507 (2007).

17    S. Kim et al., High-mobility and low-power thin-film transistors based on multilayer $MoS_2$ crystals, Nature Comm. 3, 1011 (2012).

18    S. Ghatak, A. N. Pal & A. Ghosh, Nature of electronic states in atomically thin $MoS_2$ field-effect transistors, ACS Nano 5, 7707 (2011).

19    W. Bao, X. Cai, D. Kim, K. Sridhara & M. S. Fuhrer, High mobility ambipolar $MoS_2$ field-effect transistors: substrate and dielectric effects, Appl. Phys. Lett. 102, 042104 (2013).

20    W. Park et al., Oxygen environmental and passivation effects on molybdenum disulfide field effect transistors, Nanotechnology 24, 095202 (2013).

21    B. Chakraborty, H. S. S. Ramakrishna Matte, A. K. Sood & C. N. R. Rao, Layer-dependent resonant Raman scattering of a few layer $MoS_2$, J. Raman Spectrosc. 44, 92 (2013).

22    K. Lee et al., Electrical characteristics of molybdenum disulfide flakes produced by liquid exfoliation, Adv. Mat. 23, 4178 (2011).

23    J.-H. Chen, C. Jang, S. Xiao, M. Ishigami & M. S. Fuhrer, Intrinsic and extrinsic performance limits of graphene devices on $SiO_2$, Nature Nanotech. 3, 206 (2008).

24    W. Zhu, V. Perebeinos, M. Freitag & P. Avouris, Carrier scattering, mobilities and electrostatic potential in monolayer, bilayer and trilayer graphene, Phys. Rev. B 80, 235402 (2009).

25    E. J. Kirkland, *Advanced Computing in Electron Microscopy*; Springer: New York, 2010.

26    L. P. Hansen et al., Atomic-scale edge structures on industrial-style $MoS_2$ nanocatalysts, Angew. Chem. 50, 10153 (2011).

27    H.-P. Komsa et al., Two-dimensional transition metal dichalcogenides under electron irradiation: defect production and doping, Phys. Rev. Lett. 109, 035503 (2012).

28    E. S. Kadantsev & P. Hawrylak, Electronic structure of single $MoS_2$ monolayer, Solid State Comm. 152, 909 (2012).

29    N. Tessler, Y. Preezant, N. Rappaport & Y. Roichman, Charge transport in disordered organic materials its relevance to thin-film devices: a tutorial review, Adv. Mat. 21, 2741 (2009).

30    M. Y. Han, J. C. Brant & P. Kim, Electron transport in disordered graphene nanoribbons, Phys. Rev. Lett. 104, 056801 (2010).

31    Z. G. Yu & X. Song, Variable range hopping and electrical conductivity along the DNA double helix, Phys. Rev. Lett. 86, 6018 (2001).

32    N. F. Mott, Electronic processes in non-crystalline materials, Clarendon Press, Oxford 1979.

33    F. W. Van Keuls, X. L. Hu, H. W. Jiang & A. J. Dahm, Screening of Coulomb interactions in two-dimensional variable-range hopping, Phys. Rev. B 56, 1161

(1997).

34    A. Castellanos-Gomez et al., Electric-field screening in atomically thin layers of $MoS_2$: the role of interlayer coupling, Adv. Mat. 25, 899 (2013).

35    S. Das, H.-Y. Chen, A. V. Penumacha & J. Appenzeller, High performance multilayer $MoS_2$ transistors with scandium contacts, Nano Lett. 13, 100 (2013).

36    W. Mönch, Valence-band offsets and Schottky barrier heights of layered semiconductors explained by interface-induced gap states, Appl. Phys. Lett. 72, 1899 (1998).

37    I. Popov, G. Seifert & D. Tománek, Designing electrical contacts to $MoS_2$ monolayer: a computational study, Phys. Rev. Lett. 108, 156802 (2012).

38    J. Brivio, D. T. L. Alexander, and A. Kis, Ripples and Layers in Ultrathin MoS2 Membranes, *Nano Lett.* **11**, 5148 (2011).

39    Kresse, G. & Furthmüller, J. Efficient iterative schemes for *ab* initio total-energy calculations using a plane-wave basis set. *Phys. Rev. B* **54**, 11169-11186 (1996).

40    Blöchl, P. E. Projector augmented-wave method. *Phys. Rev. B* **50**, 17953-17979 (1994).

41    Perdew, J. P., Burke, K. & Ernzerhof, M. Generalized Gradient Approximation Made Simple. *Phy.s Rev. Lett.* **77**, 3865-3868 (1996).

42    Radisavljevic, B. & Kis, A. Mobility engineering and a metal-insulator transition in monolayer $MoS_2$. Nature Mater. 12, 815-820 (2013).

43    van der Zande, A. M et al. Grains and grain boundaries in highly crystalline monolayer molybdenum disulfide. Nature Mater. 12, 554-561 (2013).

44    Najmaei, S. et al. Vapour phase growth and grain boundary structure of molybdenum disulfide atomic layers. Nature Mater. 12, 754-759 (2013).

45    S. Lebègue & O. Eriksson, Electronic structure of two-dimensional crystals from *ab initio* theory, *Phys. Rev. B* **79**, 115409 (2009).

46    W. Ren, J. Wang & Z. Ma, Conductance fluctuations and higher order moments of a disordered carbon nanotube, *Phys. Rev. B* **72**, 195407 (2005).

47    W. Ren, F. Xu & J. Wang, Emittance fluctuation of mesoscopic conductors in the presence of disorders, *Nanotechnology* **19**, 435402 (2008).

**Figure Captions**

**Figure 1 Electrical and Raman characteristics of a single-layer MoS$_2$ device.** (a) $I_{ds}$-$V_g$ characteristics of a back-gated single-layer MoS$_2$ devices in both linear and log scale under $V_{ds}$=100mV. Inset is the AFM of the device. (b) Raman spectroscopy using a 514nm laser on the device to confirm single-layer.

**Figure 2 Electron hopping transport in MoS$_2$.** (a) Schematics of electron transport mechanism in perfect (top) and defective (bottom) MoS$_2$. In perfect MoS$_2$, electron density is periodic in space and transport is band-like. In defective MoS$_2$ however, electrons are localized near the defects and transport is through hopping. (b) Arrhenius plot of normalized conductivity (symbols) of the device in Fig. 1 and the fitting results by hopping model (lines). From top to bottom, $n$=25, 20, 15, 10, 5, 1x10$^{11}$ cm$^{-2}$ respectively. The curves are offset for clarity. The two hopping regimes are clearly separated by $T^*$ (dashed vertical line). (c) $T_0$ and (d) $T_1$ obtained from the fitting of the Arrhenius plot for bare (black) and PMMA-coated (red) sample. The Arrhenius plot after PMMA coating is presented in supplementary information.

**Figure 3 TEM characterizations of MoS$_2$ showing the evidence of SV.** (a) Atomic structures of a single-layer MoS$_2$ by aberration-corrected TEM. The SV are highlighted by red arrows. Upper inset shows the MoS$_2$ sample edge to confirm the single-layer nature. Lower inset shows the schematics of the highlighted region. Scale bar, 1nm. (b) Intensity profile of along lattices with (red symbol) and without (black symbol) SV, along with simulations of a single SV (red line). The corresponding sections are highlighted in (a) by dashed lines. Inset shows the simulated TEM image of a single SV. (c) Histogram of SV density. The density was obtained by counting the number of SV in 5x5 nm$^2$ areas.

**Figure 4 DFT calculations of band structure of MoS$_2$ supercell with a single SV and the charge distribution of localized states.** (a) Band structure (left panel) and partial DOS (right panel) for single layer MoS$_2$ 5x5 supercell with a SV (see Supplementary Figure S4 for the supercell structure). The localized states are highlighted by red lines. Green dashed line corresponds to the case without SV. (b) Blue solid line, radial distribution of charge density for the localized midgap state (Band B) in (a); green dashed line, radial distribution of charge density for the delocalized top of valance band in perfect MoS$_2$. The origin is located at the SV. Inset is the isosurface ($\rho = 7 \times 10^{-3}$ e$\text{Å}^{-3}$) of the decomposed charge density corresponding to the band B in (a).

Figure 1

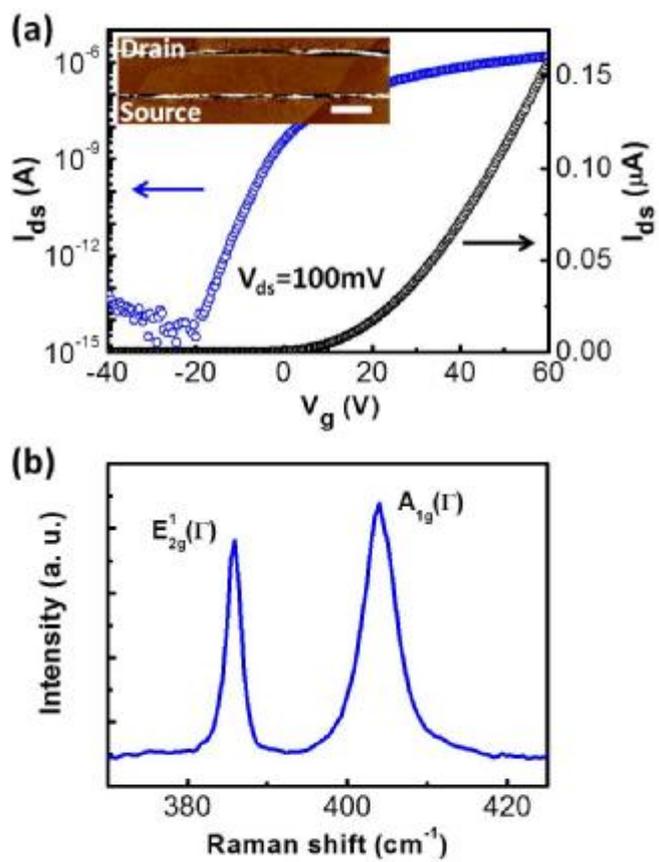

Figure 2

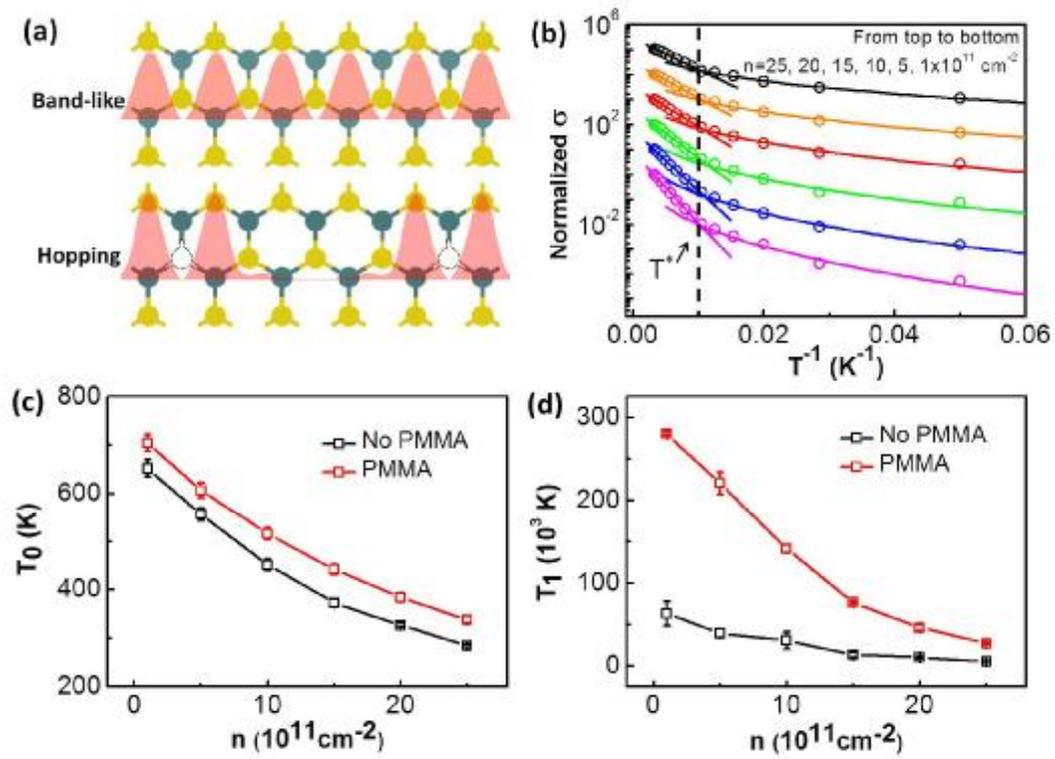

Figure 3

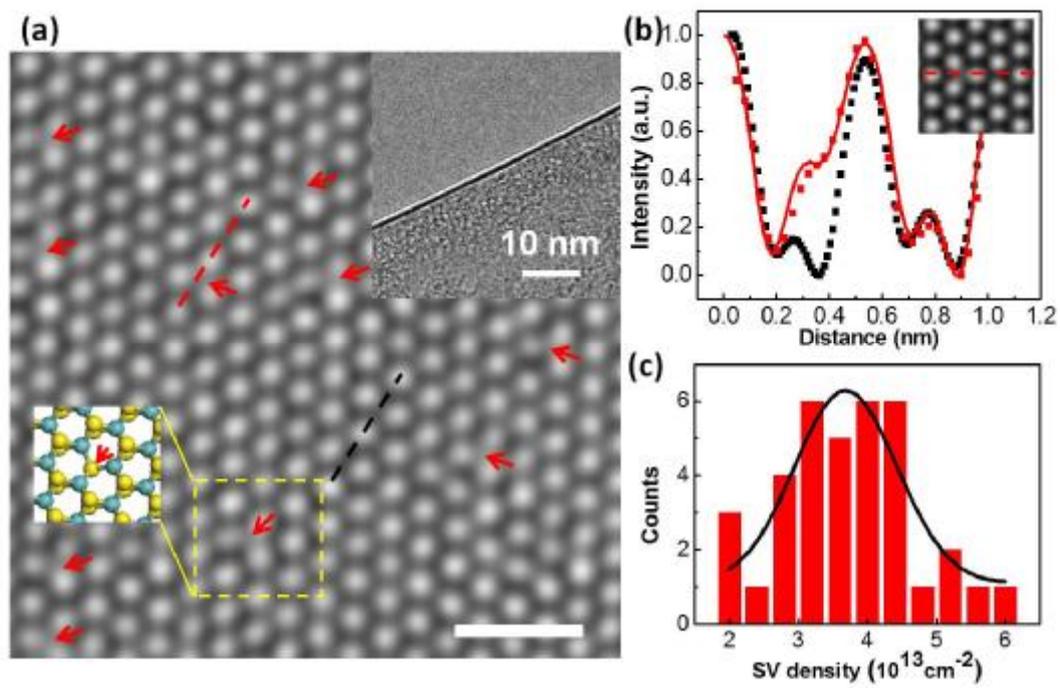

Figure 4

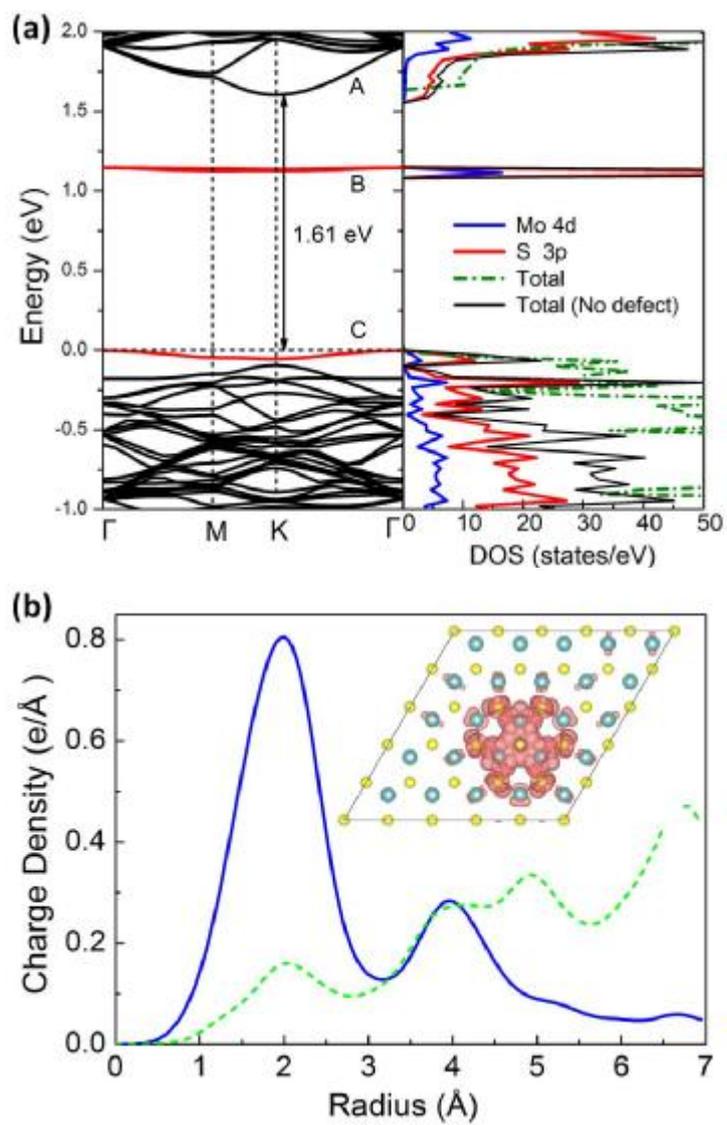

**Supplementary Figures**

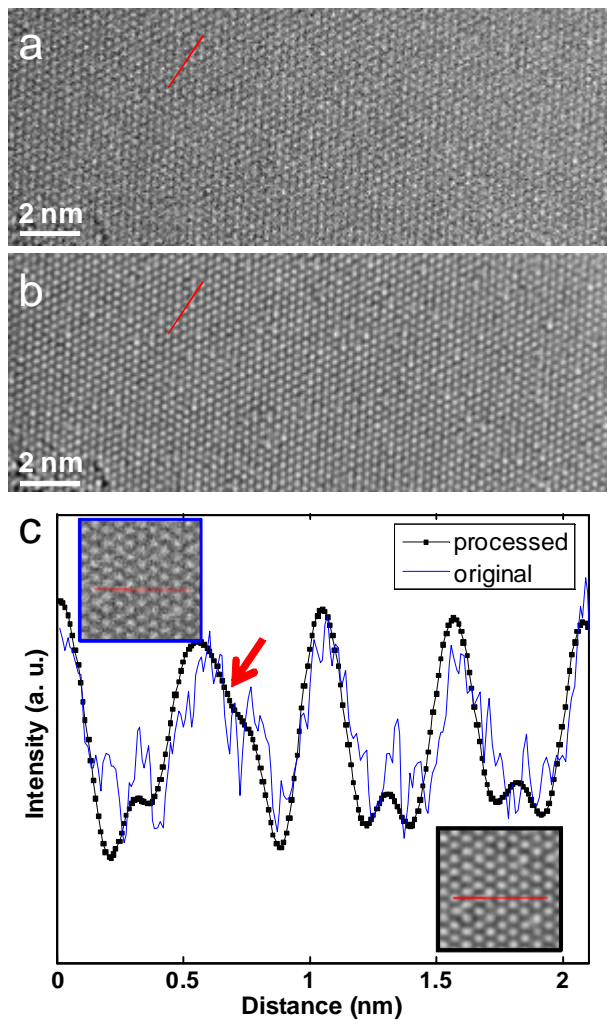

Supplementary Figure S1. (a) Original TEM image of monolayer MoS2 with SVs. (b) The same region after applying a bandpass filter to the Fast Fourier Transformation. (c) Intensity profile from original and processed TEM image (insets). The red arrow points to a single SV.

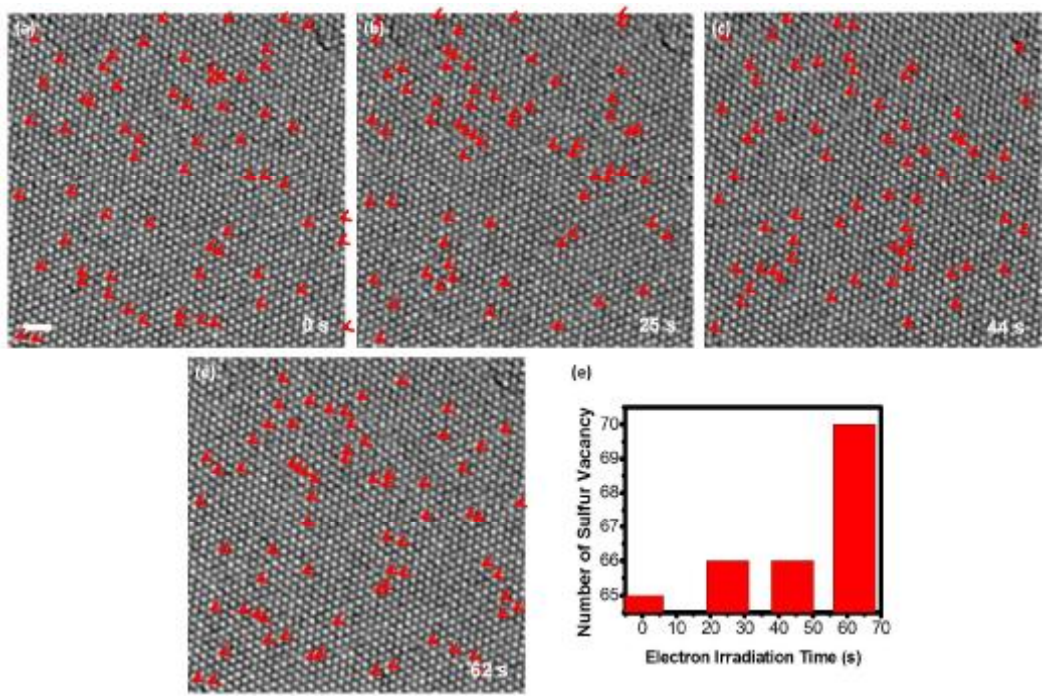

Supplementary Figure S2. (a-d) Time series of HRTEM images in the same area showing the evolution of SV in a single-layer $MoS_2$ sample. Scale bar represent 1nm. (e) Time evolution of SV number under electron irradiation.

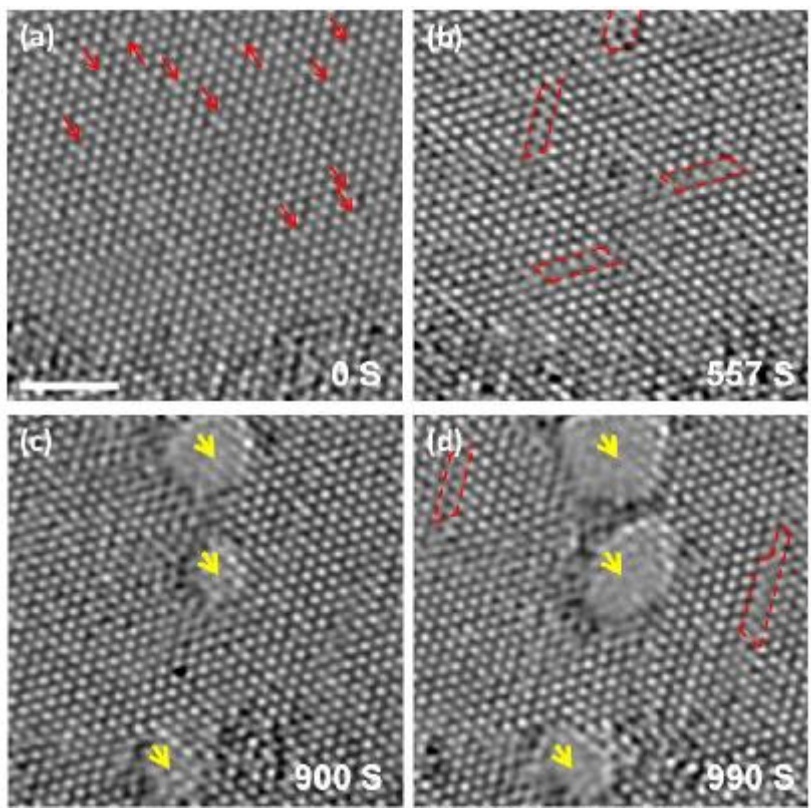

Supplementary Figure S3. Time series of HRTEM images in the same area showing the formation of complex defect in single layer of $MoS_2$. Scale bar represent 2 nm.

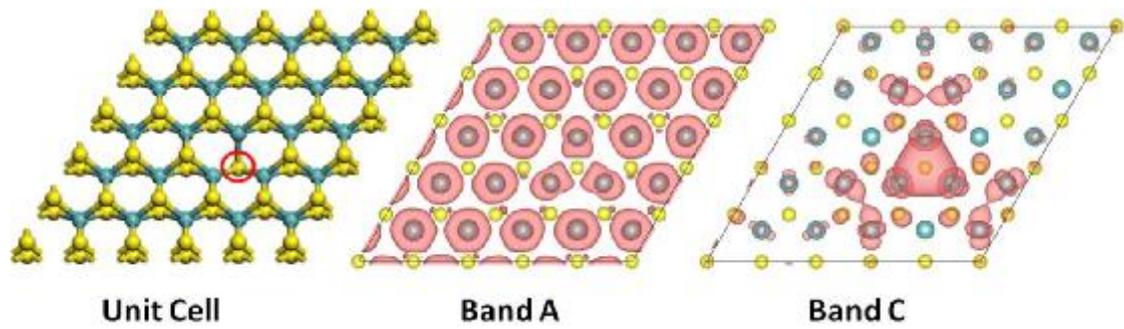

**Unit Cell**          **Band A**          **Band C**

Supplementary Figure S4. Left, the unit cellused in the calculation of Fig. 3 in the main text. The red circle is the SV. Middle and right, isosurfaces ($\rho= 7\times10^{-3}$ e/Å$^3$) of the decomposed charge density corresponding to the band A and C in Fig. 3a, respectively.

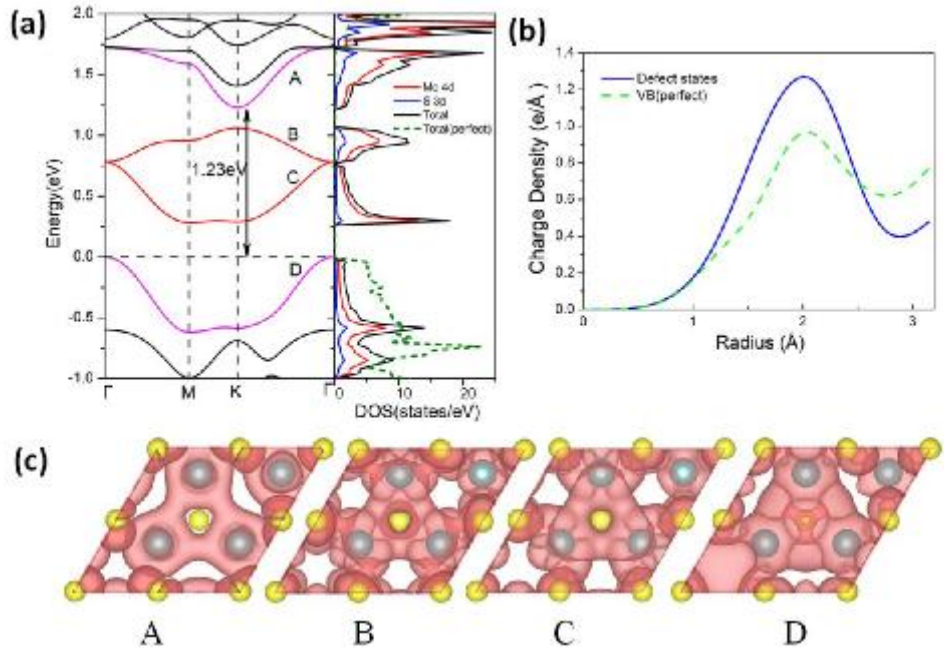

Supplementary Figure S5. (a) Band structure (left panel) and partial density of states (right panel) for 2×2 single layer $MoS_2$ with a SV. The localized states are highlighted by red lines. Green dashed line corresponds to the case without SV. (b) Blue solid line, radical distribution of charge density for the localized midgap state (Band B) in (a); green dashed line, radical distribution of charge density for the delocalized top of valance band in perfect $MoS_2$. The origin is located at the SV. (c) From left to right, isosurfaces ($\rho=4.6\times10^{-3}$ e/$\mathring{A}^3$) of the decomposed charge density corresponding to the bands labeled as A, B, C, D in (a).

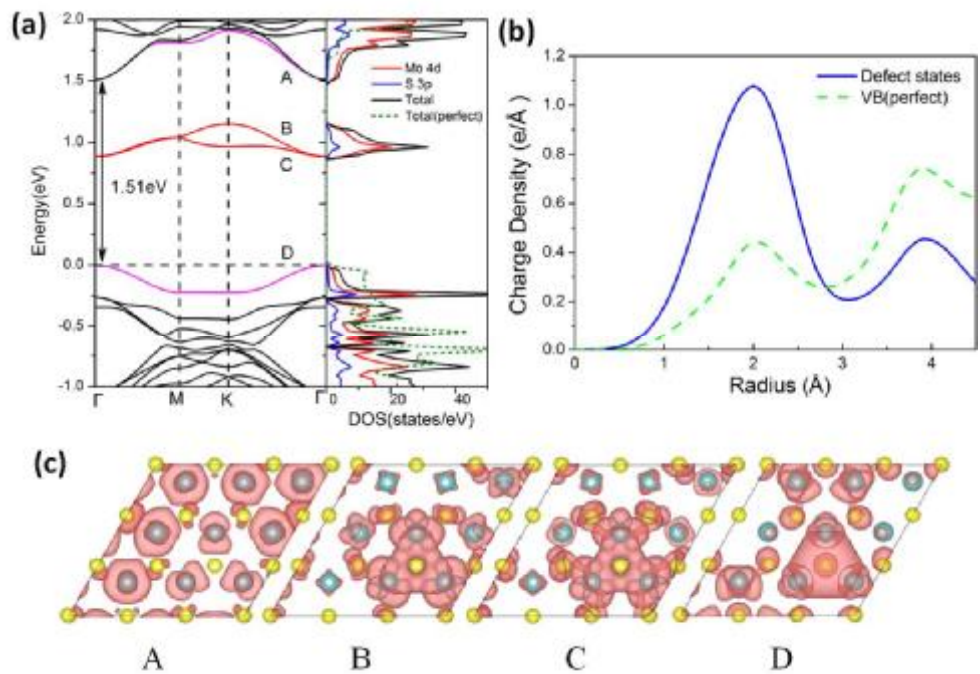

Supplementary Figure S6. Band structure, DOS, charge density of 3×3 single layer MoS$_2$ with a SV.

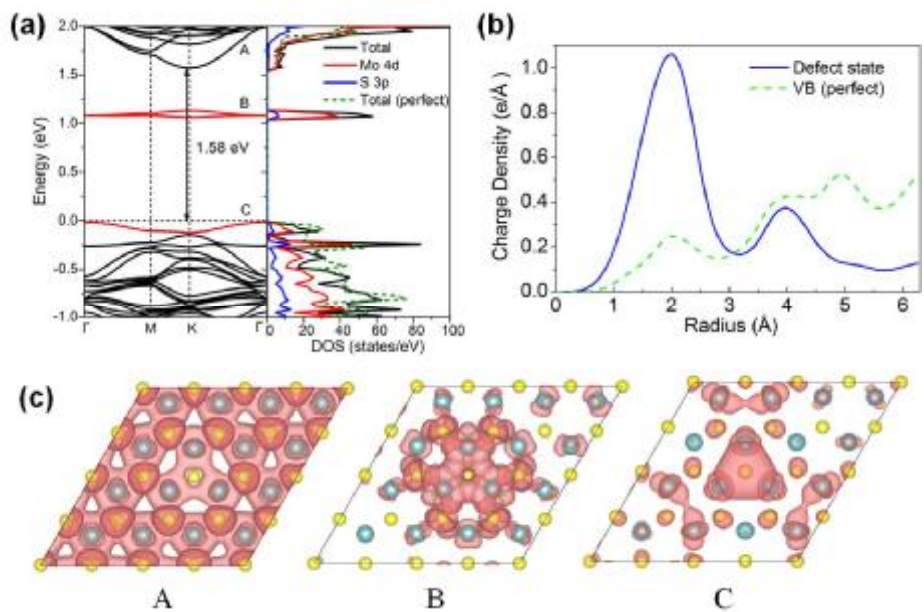

Supplementary Figure S7. Band structure, DOS, charge density of 4×4 single layer MoS$_2$ with a SV.

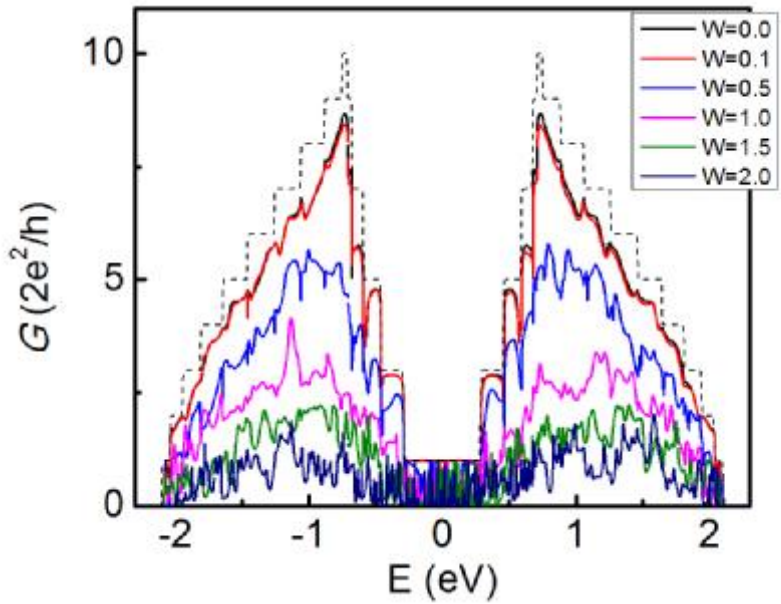

Supplementary Figure S8. Conductance as a function of Fermi energy for $MoS_2$ monolayer. The disorder strength is shown to drive the electronic transport from ballistic, diffusive, toward Anderson localized regime as $W$ increases. The total number of perfectly conducting channels is shown by the dashed staircase-like line.

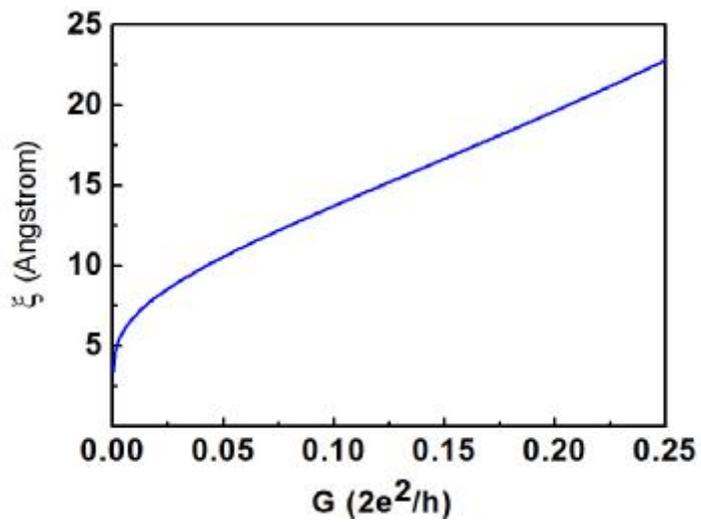

Supplementary Figure S9. Localization length $x$ as a function of the averaged

conductance $G$ extracted from the tight binding calculation and Anderson model.

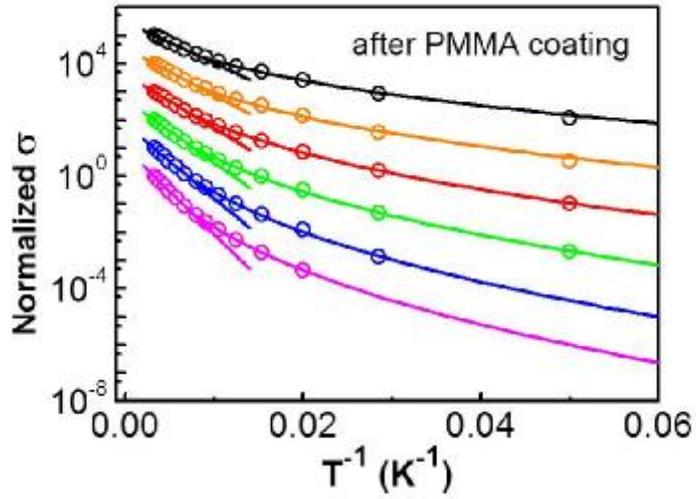

Supplementary Figure S10. Arrhenius plot of normalized conductivity (symbols) of the device in the main text after PMMA coating. Lines are the fitting results by hopping model, the fitting parameters are plotted in Fig. 2 in the main text. From top to bottom, n=25, 20, 15, 10, 5, $1\times10^{11}$ cm$^{-2}$ respectively. The curves are offset for clarity. For $n$=$1\times10^{11}$ cm$^{-2}$ and $5\times10^{11}$ cm$^{-2}$, the data points are low temperature are missing because the current is below the detection limit of our instrument.

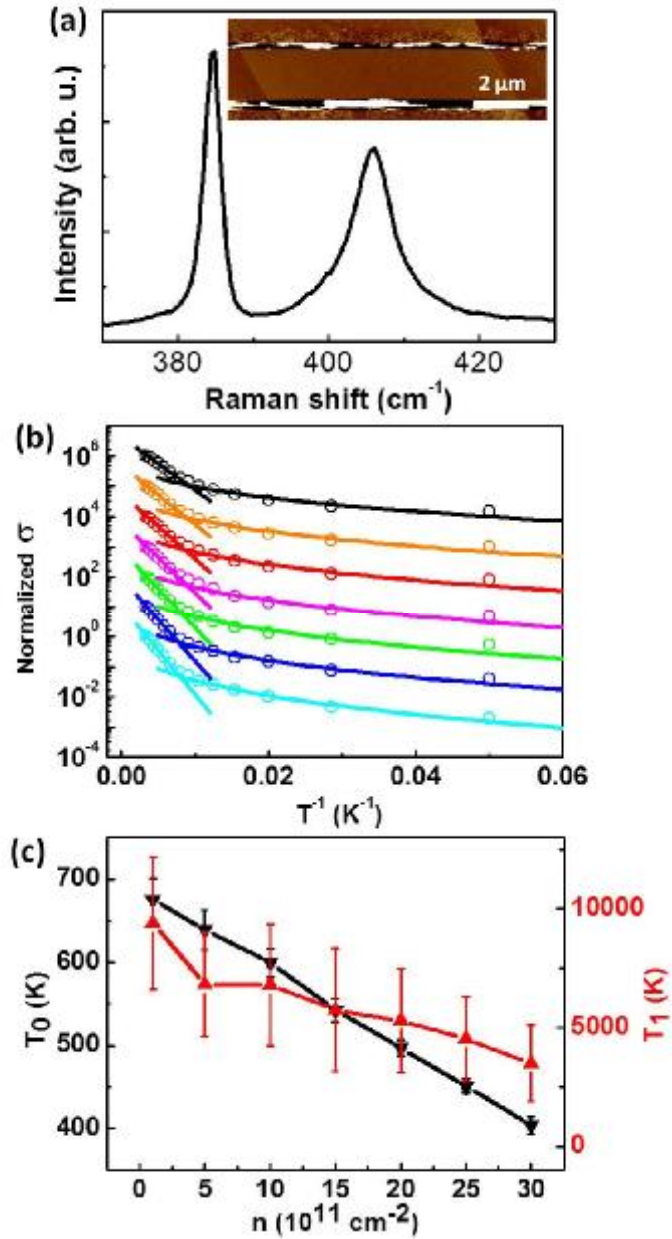

Supplementary Figure S11. Transport data on a bi-layer $MoS_2$ device. (a) Raman spectroscopy using a 514nm laser on the device to confirm bi-layer[4]. Inset shows the AFM of the device. (b) Arrhenius plot of normalized conductivity (symbols) of the device and the fitting results by hopping model (lines). From top to bottom, n=30, 25, 20, 15, 10, 5, $1x10^{11}$ $cm^{-2}$ respectively. The curves are offset for clarity. The bi-layer sample showed the same qualitative behavior as the single-layer sample in the main text. (c) $T_0$ and $T_1$ obtained from the fitting of the Arrhenius plot.

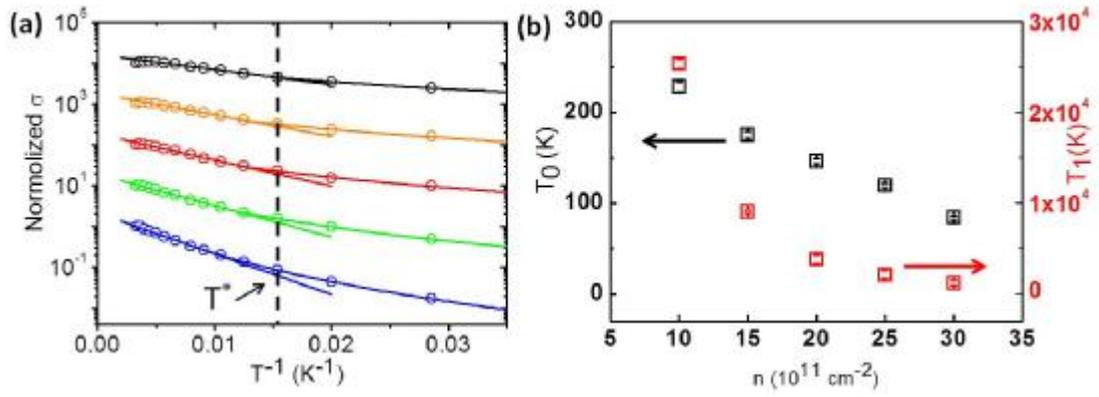

Supplementary Figure S12. (a) Arrhenius plot of normalized conductivity (symbols) of the device and the fitting results by hopping model (lines). From top to bottom, n=30, 25, 20, 15, 10x$10^{11}$ cm$^{-2}$ respectively. The curves are offset for clarity. The two hopping regimes are clearly separated by $T^* \sim$ 65K (dashed vertical line). (b) $T_0$ and $T_1$ obtained from the fitting of the Arrhenius plot.